\documentclass[reprint, prb, superscriptaddress]{revtex4-2}

\usepackage{xcolor}
\usepackage{amssymb, amsmath}
\usepackage{graphicx}
\usepackage{listings}
\usepackage{float}

\begin{document}

\title{\textit{Ab Initio} Polaron Wave Functions}
\author{Paul J. Robinson}
\affiliation{Department of Chemistry, Columbia University, New York, New York 10027, United States}

\author{Joonho Lee}
\affiliation{Department of Chemistry, Columbia University, New York, New York 10027, United States}
\affiliation{Department of Chemistry and Chemical Biology, Harvard University, Cambridge, Massachusetts 02138, United States}

\author{Ankit Mahajan}
\affiliation{Department of Chemistry, Columbia University, New York, New York 10027, United States}

\author{David R. Reichman}
\email{drr2103@columbia.edu }
\affiliation{Department of Chemistry, Columbia University, New York, New York 10027, United States}

\begin{abstract}
In this work we demonstrate that accurate ground state wave functions may be constructed for polarons in a fully {\em ab initio} setting across the wide range of couplings associated with both the large and small polaron limits.  We present a single general unitary transformation approach which encompasses an {\em ab initio} version of the Lee-Low-Pines theory at weak coupling and the coherent state Landau-Pekar framework at strong coupling while interpolating between these limits in general cases.  We show that perturbation theory around these limits may be performed in a facile manner to assess the accuracy of the approach, as well as provide an independent route to the {\em ab initio} properties of polarons.  We test these ideas on the case of LiF, where the electron-polaron is expected to be large and relatively weakly coupled, while the hole-polaron is expected to be a strongly coupled small polaron.
\end{abstract}
\maketitle

The interaction between charge carriers and phonons in solids controls a host important phenomena ranging from transport to superconductivity \cite{AshcroftNeilW1976Ssp}. 
 One of the most ubiquitous consequences of these interactions is the formation of polarons, quasiparticles composed of a charge carrier and a surrounding cloud of phonons \cite{1984Paei, franchini_polarons_2021}. Polarons can have dramatically different properties from those of bare electrons and holes.
For example, the quasiparticle band gap, carrier mass, and optical conductivity can all be strongly renormalized by polaron formation \cite{mahan2000}.

The theoretical description of polarons has evolved over the better part of the past 70 years, and has historically centered on a small number of canonical model Hamiltonians \cite{mahan2000}.  Such models represent idealized limits of the distinct physical situations in which polarons arise,  stripped from the complications associated with the detailed microscopic electronic structure of the underlying solid.
Some prominent examples include the Fr{\" o}hlich model, the Holstein model and the Su-Schrieffer-Heeger (SSH) model.  The Fr{\" o}hlich model describes the interaction of a free electron with a continuous polarizable medium \cite{Frohlich1954}, and has provided important insights into large polaron formation in polar and ionic solids such as strontium titanate (SrTiO$_{3}$), silver chloride (AgCl), and lithium fluoride (LiF)\cite{PhysRev.116.526}.  The Holstein model describes electrons on a lattice interacting with non-dispersed local phonons with a uniform on-site electron-phonon interaction (EPI)~\cite{holstein1959studiesI, holstein1959studiesII}. This model is appropriate for the study of small polarons in molecular crystals.
The (electron-phonon) SSH model describes situations where nearest neighbor lattice displacements modulate the hopping of charge between lattice sites, and has been instrumental in describing excitations such as solitons in polyenes \cite{Su1979}.
Each of these complementary models is rich in phenomenology, and the numerous studies of their solutions in different parameter regimes has provided a foundation for the understanding of polarons in a multitude of settings.  Real solids are expected to contain features of all of these models, as well as details excluded by them.

Recently, significant effort has been put forward to describe the properties of polarons starting from the {\em ab initio} Hamiltonian of a system that incorporates the full electron and phonon band structures, as well as the wave vector-dependent coupling of charge and phonons within these bands.
Among these approaches, we highlight several which have a connection to the framework we describe in this paper.  
In Ref.~\cite{Sio2019} a Landau-Pekar-like variational ansatz was developed 
and employed to study electron and hole polarons in a variety of solids.  The approach avoids the use of real space supercells by working directly within the first Brillouin zone, and provides a precise way to quantify which phonon modes contribute to specific polaronic properties.
Ref.~\cite{PhysRevMaterials.5.063805} approaches the same problem via the use of a simple canonical transformation on the {\em ab initio} Hamiltonian to efficiently recover polaron binding energies in specified limits.
Connections between these two methods have been elucidated which demonstrate that these seemingly distinct theories effectively employ the same strong coupling ansatz, and thus are appropriate for strongly coupled, highly localized polarons such as those formed in the valence band of LiF~\cite{PhysRevB.105.155132}.
More recently, a Green's function approach has been developed which provides a robust many-body framework for calculating polaron properties for {\em all} couplings~\cite{PhysRevLett.129.076402,PhysRevB.106.075119}.
Although significantly more computationally involved than the simple approaches of Refs.~\cite{PhysRevMaterials.5.063805,PhysRevB.105.155132}, this framework has the advantage of not only generality with respect to coupling strength, but also, in principle, the access to frequency-dependent quantities such as spectral functions as well as finite charge concentration effects.

Our goal in this work is to demonstrate that ground state wave functions may be constructed to accurately calculate {\em ab initio} polaronic properties for {\em arbitrary} coupling strengths in a simple and computationally efficient manner.  Specifically, we develop a canonical transformation approach which is capable of capturing {\em ab initio} binding energies of polarons in both the large polaron, weak coupling limit, as well as small polaron, strong coupling limit.  We then demonstrate that Rayleigh-Schr\"{o}dinger (RS) perturbation theory may be successfully employed to capture these limits as well, with results that show remarkable consistency for polaron binding energies with the canonical transformation approach.  These results demonstrate simple, accurate, and computationally efficient routes to the {\em ab initio} calculation of the properties of polarons in solids for arbitrary electron-phonon coupling strengths.

\begin{figure}[t]
    \centering
    \includegraphics[width=0.85\columnwidth]{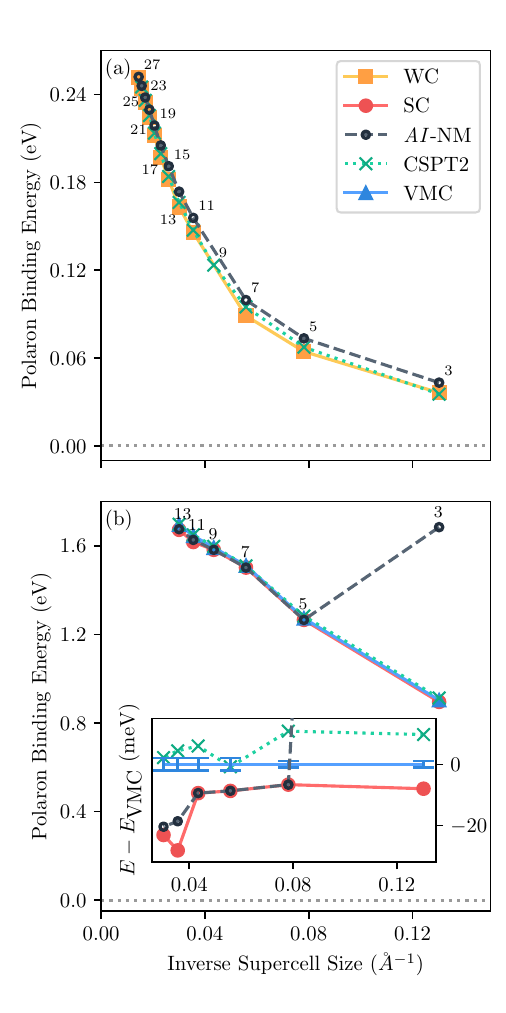}
    \caption{
    LiF electron- and hole-polaron binding energies calculated via the WC ansatz (orange squares), the SC ansatz (red circles), the \textit{AI}-NM ansatz (open black circles), CSPT2 (green Xs and dotted line), and the exact VMC approach of Ref. \cite{Mahajan2023} (blue triangles).  Each of these approaches are described in the main text and SM.
    (a) Plot of the convergence of the electron-polaron binding energy with respect to the inverse supercell size (as defined in Ref. \cite{Sio2019a}). The small numbers above the data indicate the size of the supercell, e.g. the size $N$ of the $N^{3}$ k-grid. 
    (b) Plot of the convergence of the hole-polaron binding energy with respect to inverse supercell size.
    The inset higlights the small differences between the methods.
    }
    \label{fig:nagy_LiF_el_hole}
\end{figure}

\begin{table*}
    \centering
    \begin{ruledtabular}
    \begin{tabular}{l  c c c c c}
     System & WC (eV) & SC (eV) & \textit{AI}-NM (eV) & CSPT2 (eV) & VMC (eV) \\
     \hline
     electron-polaron & $0.35$ & $0.24$ & $0.33$ & $0.35$ & -- \\
     hole-polaron & $0.30$ & $1.97$ & $1.94$ & $1.96$ & $1.97$   \\ 
    \end{tabular}
    \end{ruledtabular}
    \caption{Polaron binding energies extrapolated to the thermodynamic limit for the electron and hole polarons in LiF via the WC, SC, \textit{AI}-NM, CSPT2 methods (see text and SM), and a numerically exact VMC approach \cite{Mahajan2023}.
    The binding energies were determined by linearly extrapolating the energies to the thermodynamic limit in the reciprocal supercell volume using only the $23^3$ and $25^3$ k-grids for the electron-polaron and only the $11^3$ and $13^3$ k-grids for the electron polaron. 
    }
    \label{tab:LiF_extrapolate_binding}
\end{table*}

In the context of the Fr{\" o}hlich model, the well-known theory of Lee, Low and Pines (LLP) is applicable for weakly coupled electron-phonon systems~\cite{PhysRev.90.297}, while the adiabatic Landau-Pekar (LP) solution of the polaron problem is applicable for strongly coupled electron-phonon systems~
\cite{mahan2000, pekar1954, landau1933}.  As discussed by Huybrechts~\cite{Huybrechts1976, Huybrechts1977}, both theories can be unified from the perspective of a single unitary transformation with variational parameters that account for the displacement of the origin of the phonon modes as well as the momentum transfer associated with charge-phonon scattering.  Nagy and Marko{\v s} (NM) modified Huybrechts' ansatz by allowing the degree of momentum conservation to vary across the electronic band~\cite{Nagy}.  This modification rectifies several shortcomings of the original Huybrechts approach, and when applied to the Fr{\" o}hlich model produces results for the binding energy over a very wide range of coupling parameters that are essentially as accurate with respect to absolute error as the approach of Ref.~\cite{PhysRevLett.129.076402,PhysRevB.106.075119} as shown in the Supplemental Material (SM).  This approach forms the basis of our all-coupling {\em ab initio} wave function method, as outlined next.

Consider the general \textit{ab initio} electron-phonon Hamiltonian, written to leading linear order in the EPI as
\begin{align}
    H &= \sum_{iq} \epsilon_{iq} |i,q\rangle\langle i, q|  + \sum_{q,\nu} \omega_{\nu q} b_{\nu q}^\dagger b_{\nu q}
    \nonumber
    \\ &
    + \sum_{\substack{ij\nu \\ kq}} g^{ij}_{\nu}(k, q) |i,k+q \rangle \langle j,  k| (b_{\nu q} + b^\dagger_{\nu -q}).
\label{eq:ab_initio_hamiltonian}
\end{align}
Here, the electronic bands, phonon frequencies and, electron-phonon matrix elements are given by $\epsilon_{iq}$, $\omega_{\nu q}$ and $g^{ij}_\nu(k,q)$, respectively. The phonon creation (annihilation) operator for momentum $q$ is given by $b_{\nu q}^{\dagger}$ $(b_{\nu q})$.  Since in this work we only consider the case of a single electron (hole) in the conduction (valence) bands, we work in the projector basis for electrons (holes), thus simplifying the application of the transformation defined below.

We define a trial wave function
\begin{align}
    |
    \Psi_{t_{ik} h_{\nu q} a_{q}}
    \rangle &= U |0\rangle_{\text{ph}} \otimes
    |\psi\rangle_\text{el}.
\end{align}
The electronic portion of the wave function is given by the familiar configuration interaction (CI) singles form 
\begin{align}
    |\psi\rangle_\text{el} & =  \sum_{i k} t_{ik} |i, k\rangle,
\end{align}
where the label $i$ denotes the band index and $k$ the wave vector within the 1st Brillouin zone.  $|0\rangle_{\text{ph}}$ denotes the phonon vacuum, and the unitary transformation $U=e^S$ is defined as
\begin{align}
    S &= \sum_{\substack{i \nu \\ k q}} B_{\nu q} |i, k + a_q  q \rangle\langle i, k|,
    \label{eq:abinit_nagy_s}
    \\
    B_{\nu q} &\equiv h_{\nu q} b_{\nu q} - h^*_{\nu -q} b_{\nu -q}^\dagger.
\end{align}
In the following, we call this transformation the \textit{AI}-NM ({\em ab initio} Nagy-Marko{\v s}) approach.

There are three sets of variational parameters to be determined for this ansatz, namely the CI coefficients $\{t_{i k}\}$, the coefficients that modify electron-phonon scattering momentum $\{a_k\}$, and the variational phonon mode displacements $\{h_{\nu k}\}$. Thus, when the wave vector grid is sizable, thousands of variational parameters are used to parameterize the ground state. Note that when the set of parameters $\{a_k\}=1$, the transformation coincides with an {\em ab initio} version of the LLP transformation, while when $\{a_k\}=0$, an {\em ab initio} version of the Landau-Pekar theory results.  Importantly, we remark that in general the unitary rotation of the Hamiltonian cannot be carried out in explicit closed form except in particular limits.  Specifically, as discussed in the SM, the limits of weak coupling, strong coupling, and arbitrarily strong (electron) momentum-independent charge-phonon coupling with a parabolic band structure all yield simple closed expressions for the unitary transformation of the Hamiltonian.  Given the wide scope of these limits, the landscape of the energy functional will have many local minima, and we proceed by determining the transformation variationally, even without a rigorous minimum principle.  A similar procedure has been employed in electronic structure theory with respect to the unitary Coupled Cluster method.  In particular, our approach is equivalent to the UCC(2) method \cite{BartlettRodneyJ.1989AcaI}.

We determine the ground state energy variationally via minimization 
\begin{align}
     E_{g} \le E = \min_{t_{ik} h_{\nu q} a_q}\frac{
     \langle
     \Psi_{t_{ik} h_{\nu q} a_{q}}
     | H
     |
     \Psi_{t_{ik} h_{\nu q} a_{q}}
     \rangle
     }{
     \langle
     \Psi_{t_{ik} h_{\nu q} a_{q}}
     |
     \Psi_{t_{ik} h_{\nu q} a_{q}}
     \rangle
     },
     \label{eq:nagy_variational_min}
\end{align}
where $E_{g}$ is the true ground state of the system.
Through the second order in $h_{\nu q}$ the transformed energy $E$ is given as
\begin{align}
    \nonumber
     E  &= \frac{1}{\sum\limits_{ik} |t_{ik}|^2} 
     \Bigg\{
     \sum_{iq}\left(
        \epsilon_{iq}
        -
        \frac{1}{2}
        \sum_{\nu k} 
            |h_{\nu k}|^2 \mathcal{E}_{iqk}
        \right) |t_{iq}|^2 
        \\
        & + \nonumber
        \sum_{\nu q} \omega_{\nu q} |h_{\nu q}|^2 \sum_{iq} |t_{iq}|^2\\
        &-
        \sum_{\substack{ij\nu\\kq}} 
        \left(
            g_\nu^{ij}(k, q)
                h_{\nu -q} t_{i k+q}^* t_{j k+a_q q}
                + \text{c.c.}
        \right) \Bigg\}, 
        \label{eq:nagy_energy}
        \\
        \mathcal{E}_{iqk} & \equiv 2 \epsilon_{i q} - \epsilon_{i q - a_k k} - \epsilon_{i q + a_k k},
\end{align}

We define all of our ``off-grid" quantities via Fourier interpolation.
This is similar in spirit to Wannier interpolation, a well-known approach to bridge low density k-mesh calculations and higher density ones   \cite{PhysRevB.75.195121}.
This procedure not only provides a well-defined procedure to evaluate our energy for any set of variational parameters, but as is detailed in the SM, we can also evaluate the gradients of Eq. \ref{eq:nagy_energy} analytically, allowing efficient and direct minimization for Eq. \ref{eq:nagy_variational_min}.  A discussion of the procedure for determining the variational parameters may be found in the SM.

To illustrate the range and accuracy of our all-coupling \textit{AI}-NM approach, we consider the case of LiF where the electron-polaron is expected to behave as a relatively large polaron while the hole-polaron is expected to form a small polaron.  We work entirely within the first Brillouin zone and converge energies via a simple Makov-Payne extrapolation.  For the hole-polaron we consider only the three highest energy valence bands while for the electron-polaron we consider only the lowest energy conduction band.  Details of the {\em ab initio} calculation and parameters are contained in the SM.

In addition to the \textit{AI}-NM ansatz, one can restrict the form of Eq. 2 to the extreme limits of weak-coupling (WC), for which we set $a_k=1$ for all $k$ and strong-coupling (SC) for which we set $a_k=0$ for all $k$.  Note that for the SC case, the resulting transformation is one where coherent-state nuclear wave functions are employed as a basis. We also consider second-order RS perturbation theory in conjunction with coherent states (CSPT2) (see SM for details).  This approach is the {\em ab initio} version of the coherent-state M\o{}ller-Plesset perturbation theory formulated for the Hubbard-Holstein model previously \cite{PhysRevB.103.115123}. In the limits of very strong or very weak coupling, this method provided a second order energy correction about the Landau-Pekar or non-interacting limits, respectively. 
As discussed in the SM, it is not possible to \textit{a priori} determine whether the variational SC solution or the zero-displacement solution is the more appropriate reference for perturbation theory.
For the case of the electron- and hole-polarons we uniformly apply the zero-displacement or variational displacement solutions as the reference states respectively.
Lastly, in the case of the hole-polaron we employ the essentially exact VMC approach of Ref. \cite{Mahajan2023} as a benchmark. As we will see, we find a remarkable consistency between all of these approaches for both the electron and hole polaron systems.

In Fig. \ref{fig:nagy_LiF_el_hole} we show the grid size dependence of the electron and hole polaron binding energies for the different approaches mentioned above.  Extrapolation with respect to increasingly larger wave vector grids yields a rather consistent picture of the binding energy for both the electron and hole polarons in LiF.  These extrapolated values can be found in Table~\ref{tab:LiF_extrapolate_binding}.  The following conclusions may be reached.  With respect to the hole-polaron, a consistent binding energy close to 2 eV is found. 
We expect that this result is essentially exact within the linear electron-phonon coupling model and the level of {\em ab initio} theory used to construct the Hamiltonian. The largest uncertainty in the value of our binding energies likely arises due to the use of a crude extrapolation to the infinitely fine grid (thermodynamic) limit. The simple SC theory ($a_k=0$ for all $k$) already appears sufficient to capture this result, with almost no change in the binding energy when second order perturbation theory is performed with these states, again suggesting that convergence has been achieved.  The final value is very similar to the results for the hole-polaron found in Refs.~\cite{Sio2019, PhysRevMaterials.5.063805, PhysRevB.105.155132, PhysRevLett.129.076402,PhysRevB.106.075119}.  This is not surprising, as the SC result is equivalent to an {\em ab initio} version of Landau-Pekar theory.

\begin{figure}
    \centering
    \includegraphics[width=0.85\columnwidth]{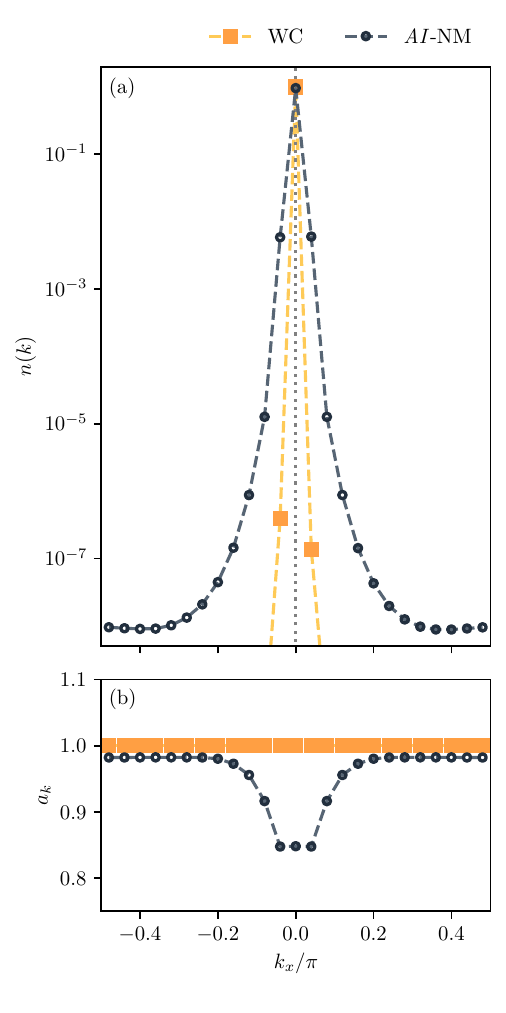}
    \caption{Detailed view of the electron-polaron results for the $25^3$ k-grid. (a) Comparison of a slice of the momentum density, $n(k) =  \langle a_k a_k^\dagger \rangle,$ for WC and {\em AI}-NM. The dotted line indicates the delta function of an infinitely localized $n(k)$ in the thermodynamic limit.  (b) The value of the momentum-conservation-modulating variational parameter $a_k$ for the $\textit{AI}-NM$ and $WC$ methods demonstrating that non-trivial $a_k$ structure can be present in large \textit{ab initio} systems.} 
    \label{fig:LiF_compare_N25}
\end{figure}

The case of the electron-polaron is more subtle.  Here we do not have VMC results to compare with.  The pure SC result gives values essentially identical to~\cite{Sio2019}.  Again this is expected as the Landau-Pekar result.  However the WC and \textit{AI}-NM approach give results consistent with each other which are close to fifty percent larger than the result found for the electron-polaron in Ref.~\cite{Sio2019}.  Remarkably, the same value is found from the CSPT2 approach, again suggesting some level of convergence and confidence in the value of the electron-polaron binding energy.  It should be pointed out that this value is smaller than that found more recently in Refs.~\cite{PhysRevLett.129.076402,PhysRevB.106.075119}. However it is difficult to compare these values as ~\cite{PhysRevLett.129.076402,PhysRevB.106.075119} include effects that arise from the second-order Debye-Waller term absent in our Hamiltonian (Eq.1) and our calculations.

It may appear that sole the utility of the full variational \textit{AI}-NM approach resides in interpolating between the WC and SC limits in an {\em ab initio} fashion.  However we note that the ansatz of Eq.2-5 is flexible enough to describe physics beyond these strict limits.  To illustrate this we note that in the pure WC theory, the momentum density distribution of the polaron is a delta function in k-space, and thus the large polaron is unrealistically delocalized.  Within the full approach of Eq.2-5, however, solutions are found that partially localize the polaron even though the binding energy is nearly identical to the WC value.  This is illustrated in Fig.\ref{fig:LiF_compare_N25}.  Localization occurs because low-lying states in the landscape of variational parameters can have non-uniform $a_k$ values between 0 and 1.  We note that a full reconstruction of the {\em observable} momentum density distribution ($n(k)=\sum_{i}n_{i}(k)$ with the sum taken over bands labels) requires undoing the unitary transformation which in fully {\em ab initio} problems necessitates an expansion in the $h_{\nu q}$ parameters which is rather involved. Thus in Fig.\ref{fig:LiF_compare_N25} only the leading-order term $n(k) \sim \sum_{i} |t_{ik}|^2$ is plotted, where the localization effect (in real-space) is already evident.

In conclusion, we have presented several routes to the {\em ab initio} calculation of the properties polarons based on ground state wave functions that encompass the full range of realistic coupling strengths in solids.  The first approach is based an all-coupling unitary transformation which naturally interpolates between known extreme limits of the problem.  Using the example of LiF, we demonstrate that this method can faithfully capture properties of both the large electron polaron and the small hole polaron, without predetermined knowledge of the physical properties of each system.  We then demonstrate that perturbation theory around the strong coupling limit provides a distinct but harmonious way to compute these properties.  Both approaches have the advantage of simplicity and appear to be highly accurate in a realistic {\em ab initio} setting.  Although not discussed here, these approaches may also form the basis of a facile means to compute real frequency spectral information as well as the properties of exciton-polarons \cite{PhysRevB.16.4480, alvertis2023phonon, PhysRevLett.132.036902, PhysRevB.109.045202, excitonUnpublished}.  These topics, as well as further calculations on a wide range of systems will be taken up in future work.

\section*{Acknowledgements}
P.J.R. acknowledges support from the National Science Foundation Graduate Research Fellowship under Grant No. DGE-2036197.
D.R.R. and A.M. are partially supported by NSF CHE-2245592.  We would like to thank Marco Bernardi and Yao Luo for helpful assistance during the early stages of this work.

\bibliography{references}

\clearpage
\appendix
\onecolumngrid

\section*{Supplemental Material}
\renewcommand{\theequation}{SM.\arabic{equation}}
\setcounter{equation}{0}
\renewcommand{\thefigure}{SM.\arabic{figure}}
\setcounter{figure}{0}

\subsection{Fr{\" o}hlich Model}

\begin{figure}[b]
    \centering
    \includegraphics{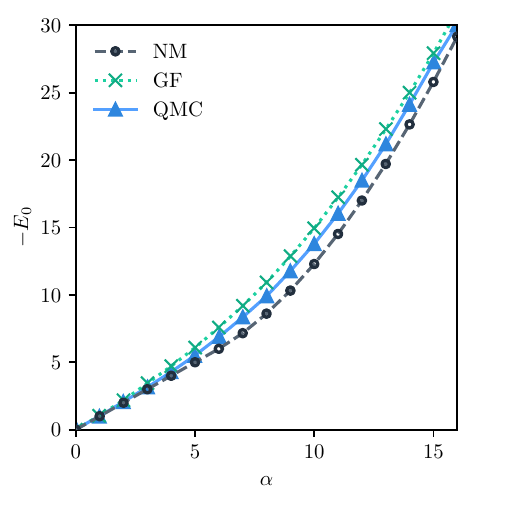}
    \caption{Exact ground state energy of the Fr{\"o}hlich model from Ref. \cite{Mishchenko2000} (blue triangles), 
    the Green's function approach of Ref. \cite{PhysRevLett.129.076402} (green X with dotted line),
    and the NM ansatz (open black circles with dashed lines).} 
    \label{fig:Nagy_frohlich}
\end{figure}

To better understand the strengths and weaknesses of the original NM ansatz, we apply it to the Fr{\" o}hlich model as was done in the original paper, and we refer the reader to Ref. \cite{Nagy} for a derivation and description.
While the main method of this work is intended for application in \textit{ab initio} systems, this section presents information about the general properties of the NM ansatz.

In Ref. \cite{Nagy} the authors considered a generalized Fr{\"o}hlich model which allows for several types of polarons to be considered.
The Hamiltonian we consider is 
\begin{align}
    H = \sum_k \epsilon_k |k\rangle \langle k| +\omega_0 \sum_q  b_q^\dagger b_q
    + \sum_{kq} g(q) |k+q\rangle \langle k | (b_q + b^\dagger_{-q}).
    \label{eq:frohlich_hamiltonian}
\end{align}
We restrict our discussion here to a dispersionless longitudinal optical (LO) phonon mode of frequency $\omega_0$ and an EPI with the form $|g(q)|^2 = \frac{4\pi \alpha \omega_0^{3/2}}{\sqrt{2 V} q^{2}}$ where $\alpha$ is the dimensionless coupling constant.
The free-particle dispersion is given by $\epsilon_k = k^2 / 2m$.

Nagy and Marko{\v s} employ a unitary transformation defined in Eq. 6 of Ref. \cite{Nagy} with the projection of the density written in the form $\sum_k |k+a_q q\rangle \langle k|$, where $a_q$ is a variational parameter defined for all $q$ to be between zero and one.  Nagy and Marko{\v s} additionally consider an ansatz for the electronic wave function of the Gaussian form (an assumption which we relax in our \textit{ab initio} version), with the k-space wave function given by $\sqrt{\frac{8}{V}}\left(\frac{\pi}{\lambda}\right)^{3/4} 
    e^{\frac{-\mathbf{k}^2}{2\lambda}}$. 
Using their transformation and this polaronic wave function one may write the energy as an integral equation dependent on $\lambda$.

The results of the NM ansatz for the Fr{\"o}hlich model are summarized in Fig. \ref{fig:Nagy_frohlich}.
As mentioned previously, the Fr{\" o}hlich model is challenging for variational approaches, and most incorrectly \cite{RevModPhys.63.63} predict a phase transition when transitioning from WC to SC in the form of a discontinuity in the energy. 
The NM ansatz predicts only a discontinuity in the derivative of the energy at $\alpha=6$ which is an improvement over a discontinuity in the energy's value.
The solution matches the perturbative WC result $(E = -\alpha)$ up to $\alpha=6$ where it then transitions not to the SC solution $(E = -\alpha^2 / 3\pi)$, 
but to a solution that asymptotically approaches the SC solution \cite{mahan2000}.
Compared to the numerically exact results of Ref. 
\cite{Mishchenko2000}, the NM ansatz never errs by more than $\approx 14 \%$ over the full range of couplings.
The Green's function approach of Ref. \cite{PhysRevLett.129.076402} has a slightly lower maximum error of $\approx 10 \%$; however, it consistently overshoots the exact energy (the approach is not variational) and does not do as well as the NM ansatz or the pure WC result for very weak couplings.
The added flexibility of the NM ansatz over both the SC and WC ansatzes allows it to closely match the exact result across all couplings. 

\subsection{{Ab Initio} Calculations}
We follow the procedure for generating $\epsilon_{ik}$, $\omega_{\nu q}$ and $g_\nu^{ij}(k, q)$ for LiF discussed in Ref. \cite{PhysRevB.106.075119}.
We begin by optimizing the geometry of the system using Quantum Espresso \cite{Giannozzi_2009, Giannozzi_2017} employing a 12x12x12 k-grid using the the PBE exchange-correlation functional \cite{PhysRevLett.77.3865} with Optimized Norm-Conserving Vanderbilt Pseudopotentials \cite{PhysRevB.88.085117,SCHLIPF201536}, and a 150 Ha energy cutoff.
We find an optimized lattice constant of $a_0 = 4.035 $ \AA.
We then apply density functional perturbation theory (DFPT) as implemented in the PHonon code within Quantum Espresso to find the phonon frequencies and electron-phonon couplings on an 11x11x11 grid.
We next construct the maximally localized Wannier functions with Wannier90 \cite{wannier90} and use EPW \cite{PONCE2016116} to interpolate the parameters onto all of the k-grids applied in this work. 

\subsection{Closure of All-Coupling Ansatz\label{app:nagy_closure}}

While the unitary transformation defined by Eq. \ref{eq:abinit_nagy_s} has appealing properties, when combined with the \textit{ab initio} Hamiltonian in Eq. \ref{eq:ab_initio_hamiltonian} it does not close in general.
We will instead work with a truncated version of the transformation
which discards all terms greater than quadratic in $h_{nq}$.  Given this truncation we examine the magnitudes of $h_{nq}$ after optimization and discard solutions with large $h_{nq}$
\footnote{
To preference against solutions which fall outside our acceptable bounds for $h_{\nu q}$, we set a cutoff $h_\text{cut}$ and then apply a forcing function to our energy of the form
$
    f(\{ h_{\nu q} \})=
    \sum_{\nu q}
        \exp{\left(c 
        (2 |h_{\nu q}|^2  - h_\text{cut}^2 )/
        h_\text{cut}^2
        \right)
        } - \exp\left(c \right)
$
if $|h_{\nu q}| > h_\text{cut}$ and zero otherwise.
$c$ is the strength of the forcing term.
}.

The transformed Hamiltonian can be expressed using the Baker–Campbell–Hausdorff (BCH) formula
\begin{equation}
    e^{-S} O e^S = O + [O, S] + \frac{1}{2!}[[O, S], S] + \dots.
    \label{eq:bch_expansion}
\end{equation}

The phonon creation/annihilation operators are readily transformed and become
\begin{align}
    e^{-S} b_{\nu q} e^{S} &= b_{\nu q} - h^*_{\nu q} \sum_{i k} | i, k \rangle\langle i, k - a_q q|, \\
    e^{-S} b_{\nu q}^\dagger e^{S} &= b_{\nu q}^\dagger - h_{\nu q} \sum_{i k} |i, k - a_q q \rangle\langle i, k|.
\end{align}

For the band term, we can write out the full transformation analytically.
\begin{align}
    H_{e} &= \sum_{i k} \epsilon_{i k} |i, k\rangle \langle i, k| \\
    e^{-S} H_{e} e^S
     &= 
     \sum_{i, k}
     \sum_{n=0}^\infty
     \frac{1}{n!}
     \sum_{\substack{
     q^{(1)}\dots q^{(n)} \\ 
     \nu^{(1)}\dots \nu^{(n)}
     }}
\left[ \prod_{m=1}^{n} B_{\nu^{(m)} q^{(m)}} \right]
C^{(n)}_{i q^{(1)} \dots q^{(n)}}(k) 
\bigg|i k \bigg\rangle\bigg\langle i k - \sum\limits_{m=1}^n a_{q^{(m)}} q^{(m)} \bigg|
\label{eq:full_band_transform}
\end{align}
Here, the coefficients are given by the recursion relation
\begin{align}
    C^{(0)}_{i}(k) &= \epsilon_{i k}, \\
    C^{(n)}_{iq^{(1)}\dots q^{(n)}}(k) &= 
    C^{(n - 1)}_{i q^{(1)}\dots q^{(n-1)}}(k) -
    C^{(n - 1)}_{i q^{(1)} \dots  q^{(n-1)}}(k-a_{q^{(n)}}q^{(n)}).
    \label{eq:band_recurrence_realation}
\end{align}
Evaluating Eq. \ref{eq:full_band_transform} in it's entirety is impractical and we will use only up to second order.
However, the structure of Eq. \ref{eq:band_recurrence_realation} demonstrates that if
$\epsilon_{i k}$ is an $m^\text{th}$-order polynomial, 
then all $C^{(n>m)} = 0$.

Considering the case of the Fr{\" o}hlich model, which has parabolic bands, we now understand why the band term for the original ansatz closed exactly on that model \cite{Nagy}.
As an example, the three non-vanishing coefficients for the Fr{\" o}hlich model are
\begin{align}
    C^{(0)}_k  &= \frac{k^2}{2 m} \\
    C^{(1)}_{k,q}  &=  -\frac{ a_q q \left(a_q q -2 k\right)}{2 m} \\
    C^{(2)}_{k, q_1, q_2} &= \frac{a_{q_1} q_1 a_{q_2}  q_2}{m}.
\end{align}
It is clear that all higher order terms will vanish because $C^{(2)}$ has no $k$ dependence, so further application of Eq.
\ref{eq:band_recurrence_realation} will produce zero.

We now consider the electron-phonon interaction term
\begin{align}
    H_\text{e-ph} = 
    \sum_{\substack{ij\nu \\ kq}} g^{ij}_{\nu}(k, q) 
    |i,k+q \rangle \langle j,  k| (b_{\nu q} + b^\dagger_{\nu -q}).
    \label{eq:interaction_ab_initio}
\end{align}
The phonon operators are easily transformed so we need only consider transforming a single momentum sum over the "density" 
$\Tilde{\rho}_{q}^{ij} = \sum\limits_k g_\nu^{ij}(k,q) |i k+q \rangle \langle j k|. $
As before we can write a recursive expression for the full transformation
\begin{align}
    e^{-S} \Tilde{\rho}_{q}^{ij} e^S
     &= 
     \sum_{n=0}^\infty
     \frac{1}{n!}
     \sum_{\substack{
     k \\
     q^{(1)}\dots q^{(n)} \\ 
     \nu^{(1)}\dots \nu^{(n)}
     }}
\left[ \prod_{m=1}^{n} B_{\nu^{(m)} q^{(m)}} \right]
D^{(n)}_{i j \nu q q^{(1)} \dots q^{(n)}} (k)
\bigg|i k + q + \sum\limits_{m=1}^n a_{q^{(m)}} q^{(m)} \bigg\rangle\bigg\langle i k  \bigg|.
\label{eq:full_density_transform}
\end{align}
Similarly to Eq. \ref{eq:band_recurrence_realation}, the coefficients are given by the recursion relation
\begin{align}
    D^{(0)}_{ij \nu q}(k) & = g_{\nu}^{ij}(k, q), \\
    D^{(n)}_{ij\nu q q^{(1)}\dots q^{(n)}}(k) &= 
    D^{(n-1)}_{ij\nu  q q^{(1)}\dots q^{(n-1)}} \left(k + a_{q^{(n)}} q^{(n)}\right)-
    D^{(n-1)}_{ij\nu q q^{(1)}\dots q^{(n-1)}}(k). 
    \label{eq:density_recurrence_realation}
\end{align}

Because we only consider terms which do not vanish upon application of the phonon vacuum and are overall second order or lower in the phonon displacement, 
our energy expression only requires the zeroth- and first-order terms in Eq. \ref{eq:full_density_transform}.
We also rearrange the first-order term so that the 
electron-phonon coupling elements are easily evaluated
\begin{align}
    e^{-S} \Tilde{\rho}_{q}^{ij} e^S 
    \approx 
    \Tilde{\rho}_{q}^{ij}
    + \sum_{\substack{k \nu' q'}} B_{\nu' q'} g_\nu^{ij}(k, q)
\left(
|i, k + q \rangle \langle j, k - a_{q'} q'|
-
|i, k + q + a_{q'} q'\rangle \langle j, k|
\right).
\label{eq:interaction_first_order}
\end{align}

As in Eq. \ref{eq:band_recurrence_realation}, if $g_{\nu}^{ij}(k, q)$ is a polynomial in $k$ with degree $m$ then all terms of order $n > m$  in Eq. \ref{eq:full_density_transform} will vanish and the transformation closes. 
In Ref. \cite{Nagy} the interaction term closed without approximation for the Fr{\" o}hlich model, which can now be understood because in that model 
the coupling is not $k$-dependent and only depends on the momentum of the emitted/absorbed phonon. 
It is important to note that this term also closes in the strong coupling limit where, as for the band term, the first- and all higher-order terms vanishes. 

Overall, there are three cases where the transformation closes and thus the method is rigorously variational.
These are the trivial weak-coupling limit ($g(k,q) \rightarrow 0$), the strong coupling ($a_k=0$) limit, and, less trivially, cases which include the Fr{\"o}hlich model where the electron-phonon coupling is independent of the electronic-momentum and the electronic bands are polynomials up to second-degree.

\subsection{Numerical Evaluation and Optimization of \textit{AI}-NM Ansatz \label{app:Nagy_numerics}}
In order to evaluate Eq. \ref{eq:nagy_energy} we must be more precise about what it means to have a continuous-valued variational parameter in the index of another parameter.
It would seem at first that this type of parameter would only make sense in the thermodynamic limit when the parameters become functions with continuous domains. 

Nonetheless, we can define the method for finite-sized systems in an unambiguous manner via the Fourier transform.
For some parameter $\eta$ which is defined in the discrete reciprocal lattice and two vectors $k$ and $q$ which are on the lattice, we define
\begin{align}
    \eta_{k + a_q q} &= \sum_r \eta_r e^{-i a_q q \cdot r} e^{-i k \cdot r} = 
    \mathcal{F}\left[\eta_r e^{-i a_q q \cdot r}; r, k\right], \\
    \eta_r &= \mathcal{F}^{-1}[\eta_k].
\end{align}
The trade-off for using the $a_k$ parameters is that now the ansatz is in a mixed momentum-position representation and requires Fourier transformations at each step.  The Fourier interpolation allows us to have a well-defined energy expression and prescribes a straightforward procedure to compute the gradients.  We note that for our optimization it is numerically more convenient for all of our parameters to have domains of $(-\infty, \infty)$, so we define $a_k = S(z_k)$ where $S(x) = (1 + \exp(-x))^{-1}$ is the Sigmoid function.

We optimize the variational parameters via L-BFGS-B minimization of Eq. \ref{eq:nagy_variational_min} \cite{2020SciPy-NMeth}.
In the equations below, we display the gradients of the energy with respect to the variational parameters. We only show the gradients of the numerator of Eq. \ref{eq:nagy_variational_min} which we denote as $E_0$. The full gradient is then constructed via standard calculus.  

Starting with the displacements
\begin{align}
    \frac{\partial E_0}{\partial h_{\nu p}^*} &=
        -\frac{1}{2}h_{\nu p} \sum_{i q} \mathcal{E}_{iqp} |t_{iq}|^2
        + \omega_{\nu p} h_{\nu p} \sum_{ik}|t_{ik}|^2
        -
        \sum_{ijk} g_{\nu}^{ij}(k, p) t_{i k+ p(1-a_p)}^* t_{jk}
\end{align}
In the Fourier representation of the problem we can show that our modified band
energy can be evaluated as 
\begin{align}
\mathcal{E}_{iqk} = 4 \sum_{r} \epsilon_{ir} \sin^2\left(\frac{k a_k \cdot r}{2}\right) e^{-i q\cdot r},
\end{align}
For the CI amplitudes
\begin{align}
    \frac{\partial E_0}{\partial t_{i p}^*} =
        &\left(
        \epsilon_{ip}
        -
        \frac{1}{2}
        \sum_{\nu k} 
            |h_{\nu k}|^2 \mathcal{E}_{ipk}
        \right) t_{ip}
        \nonumber
    \\ -
        \sum_{\substack{j\nu \\ q}}
        &
        \left(
                g_\nu^{ij}(p-q, q)
                h_{\nu -q} 
            t_{j p - q(1-a_q)}
                +
                \frac{1}{N} 
                \sum_{k r}
                g_\nu^{ij}(k, q)^*
                h_{\nu -q}^* 
                t_{i k+q}
                e^{i(k - p + a_q q)\cdot r} 
            \right).
\end{align}
The gradient with respect to $a_k$ is given by
\begin{align}
    \frac{\partial E_0}{\partial a_p} &=
        -\sum_{q i \nu}  |t_{iq}|^2 
        \left(|h_{\nu p}|^2 + |h_{\nu -p}|^2\right)
            \sum_r \epsilon_{i r} ( p\cdot r ) \sin(a_p p \cdot r) e^{-i q \cdot r} \nonumber 
            \\
        &+2 \Re\left( i \sum_{\substack{ij\nu \\ k}} g_\nu^{ij}(k, p) h_{\nu -q} t_{i k+p}^* 
            \sum_r t_{jr} (p \cdot r) e^{-i a_p p \cdot r} e^{-i k \cdot r} \right) 
        \nonumber
        \\
        &-2 \Re\left( i \sum_{\substack{ij\nu \\ k}} g_\nu^{ij}(k, -p) h_{\nu q} t_{i k-p}^* 
            \sum_r t_{jr} (p \cdot r) e^{i a_p p \cdot r} e^{-i k \cdot r} \right).
\end{align}

The application of analytic gradients allows us to efficiently run the L-BFGS-B routine; however, the optimization itself is non-trivial and presents several challenges. We found that there were multiple competing solutions, so the solver could, for example, optimize to the SC state even when there was a lower minimum available. Because of this difficulty, we ran the minimization procedures from a number of starting points including some randomly generated. Additionally, we interpolated the lowest energy solutions from smaller k-mesh sizes to larger k-meshes in order to speed up convergence. We also note a numerical error that arises in small systems because the computed values of $g_\nu^{ij}(k,q)$ are not precisely Hermitian. This non-physicality results in a slight disagreement between the evaluated energies and the gradient minimization because Hermiticity of the electron-phonon coupling is analytically assumed in our ansatz. Finally, the optimization via the approach outlined above with the minimizer we have utilized often terminates where notable gradient structure is still present.  This may be a consequence of our interpolation procedures, but further investigation is needed to improve the optimization, ensuring the lowest lying solutions are obtained.

\subsection{CSPT2}
CSPT2 is a version of MP2 which applies coherent state displacements to the electron-phonon Hamiltonian in order to generate the reference Hamiltonian for perturbation theory \cite{PhysRevB.103.115123}.
Beginning with Eq.~\ref{eq:ab_initio_hamiltonian} we apply the coherent state transformation to displace each phonon operator by a complex-scalar displacement $\phi_{\nu_\mathbf{q}}$. 
Unlike in previous sections, we consider both the variationally minimized non-zero $\phi_{\nu_\mathbf{q}}$ and the zero-displacement, $\phi_{\nu_\mathbf{q}} = 0$, solutions.
This additional consideration is important because it is not possible to know which solution is the best representation of the true wavefunction \textit{a priori}, so we use the perturbative energy to determine the most appropriate reference \textit{a posteriori}.

Beginning with Eq.~\ref{eq:ab_initio_hamiltonian} and applying the coherent state transformation, we define the zeroth-order Hamiltonian as
\begin{equation}
\hat{H}_0
=
\sum_{\mathbf k m} \epsilon_{m_{\mathbf k}} a_{m_{\mathbf k}}^\dagger a_{m_{\mathbf k}}
+
\sum_{\mathbf k \mathbf q m n \nu}
g^{nm}_\nu(\mathbf k, \mathbf q)
a_{n_{\mathbf k + \mathbf q}}^\dagger
a_{m_{\mathbf k}}
(\phi_{\nu_{\mathbf q}} + \phi_{\nu_{-\mathbf q}}^*)
+\sum_{\mathbf q \nu}
\omega_{\nu_{\mathbf q}}
|\phi_{\nu_{\mathbf q}}|^2
+\sum_{\mathbf q \nu}
\omega_{\nu_{\mathbf q}}
b_{\nu_{\mathbf q}}^\dagger
b_{\nu_{\mathbf q}}
\label{eq:h0_cspt2}
\end{equation}
The ground state of Eq.~\ref{eq:h0_cspt2} is the same as the optimized SC state,
\begin{equation}
|\Psi_0^{(0)}\rangle
=
\sum_{\mathbf km} t_{m_{\mathbf k}}
|m_{\mathbf k}\rangle |0\rangle.
\end{equation}
The excited states of Eq.~\ref{eq:h0_cspt2} can be labeled by electronic state index $\alpha$ and the phonon occupation string $\{n_{\nu_{\mathbf q}}\}$.
The corresponding electronic
states can be obtained by diagonalizing the Fock operator, 
\begin{equation}
\hat{F}
=
\sum_{\mathbf k m} \epsilon_{m_{\mathbf k}} a_{m_{\mathbf k}}^\dagger a_{m_{\mathbf k}}
+
\sum_{\mathbf k \mathbf q m n \nu}
g^{nm}_\nu(\mathbf k, \mathbf q)
a_{n_{\mathbf k + \mathbf q}}^\dagger
a_{m_{\mathbf k}}
(\phi_{\nu_{\mathbf q}} + \phi_{\nu_{-\mathbf q}}^*).
\label{eq:fock_operator_scpt2}
\end{equation}
We label eigenvalues of this Fock operator by $\epsilon_{\alpha}$ with the corresponding eigenvectors $\mathbf t_{\alpha}$.

In general, the exact ground state can be written using the basis of these states,
\begin{equation}
|\Psi_0\rangle
=
\sum_{\alpha, \{n_{\nu_{\mathbf q}}\}}
C_{\alpha,\{n_{\nu_{\mathbf q}}\}}
|\psi_\alpha\rangle |\{n_{\nu_{\mathbf q}}\}\rangle.
\end{equation}
In CSPT2, we obtain 
$C_{\alpha,\{n_{\nu_{\mathbf q}}\}}$
by perturbative expansion.
Our perturbing Hamiltonian follows
\begin{equation}
\hat{V}
=
\sum_{\mathbf k \mathbf q nm\nu}
g^{nm}_\nu(\mathbf k, \mathbf q)
a_{n_{\mathbf k + \mathbf q}}^\dagger
a_{m_{\mathbf k}}
(
b_{\nu_{\mathbf q}}
+
b_{\nu_{-\mathbf q}}^\dagger
)
+
\sum_{\mathbf q \nu}
\omega_{\nu_{\mathbf q}}
(b_{\nu_{\mathbf q}}^\dagger
\phi_{\nu_{\mathbf q}}
+
\phi_{\nu_{\mathbf q}}^*
b_{\nu_{\mathbf q}}
).
\end{equation}
It is immediately clear that
the first-order energy contribution, $E^{(1)}$, is zero.
The first-order wave function is
determined by (observing only one phonon states survive)
\begin{equation}
C_{\alpha,1_{\nu_{\mathbf q}}}^{(1)}
=
\frac{
\langle \Psi_{\alpha,1_{\nu_{\mathbf q}}}
|
\hat{V}
|
\Psi_0^{(0)}
\rangle
}{E_0^{(0)} - E_{\alpha}^{(0)}}
=
\frac{
\sum_{\mathbf k nm}
g^{nm}_\nu(\mathbf k, \mathbf q)
(t_{\alpha, n_{\mathbf k + \mathbf q}})^*
t_{0,m_{\mathbf k}}
+\delta_{\alpha,0}
\omega_{\nu_{\mathbf q}}
\phi_{\nu_{\mathbf q}}
}
{
\epsilon_0-\epsilon_{\alpha}-\omega_{\nu_{\mathbf q}}
}.
\end{equation}
Using this, 
the second-order energy contribution is
\begin{equation}
E^{(2)}
=
\sum_{\alpha \nu_{\mathbf q}}
\frac{
|\langle \Psi_{\alpha,1_{\nu_{\mathbf q}}}
|
\hat{V}
|
\Psi_0^{(0)}
\rangle|^2
}{E_0^{(0)} - E_{\alpha}^{(0)}}
=
\sum_{\alpha \nu_{\mathbf q}}
\frac{
|\sum_{\mathbf k nm}
g^{nm}_\nu(\mathbf k, \mathbf q)
(t_{\alpha, n_{\mathbf k + \mathbf q}})^*
t_{0,m_{\mathbf k}}
+\delta_{\alpha,0}
\omega_{\nu_{\mathbf q}}
\phi_{\nu_{\mathbf q}}
|^2
}
{
\epsilon_0-\epsilon_{\alpha}-\omega_{\nu_{\mathbf q}}
}.
\end{equation}
To generate the energies presented in the main text the following protocol is used. First we evaluate the energy expectation value of Eq.~\ref{eq:fock_operator_scpt2} with both $\phi_{\nu_\mathbf{q}} = 0$ and a $\phi_{\nu_\mathbf{q}}$ variationally minimized from a small random displacement.

We can also compute other properties than energy order-by-order.
In particular, we focus on
evaluating the expectation value for
\begin{equation}
\rho_n(\mathbf k) =
a_{n_{\mathbf k }}^\dagger
a_{n_{\mathbf k}}.
\end{equation}
The first order correction is
\begin{equation}
\langle \Psi^{(1)}_0
|
\rho_n(\mathbf k)
|\Psi^{(0)}_0\rangle 
+ h. c. = 0.
\end{equation}
The second order correction is
\begin{equation}
(\langle \Psi^{(2)}_0
|
\rho_n(\mathbf k)
|\Psi^{(0)}_0\rangle 
+ h.c.)
+
\langle \Psi^{(1)}_0
|
\rho_n(\mathbf q)
|\Psi^{(1)}_0\rangle.
\end{equation}
We only compute the last term for computational efficiency reasons.
This term is given as
\begin{align}
\nonumber
\langle \Psi^{(1)}_0
|
\rho_n(\mathbf k)
|\Psi^{(1)}_0\rangle 
&=
\sum_{\alpha\beta\nu_{\mathbf q'}
\lambda_{\mathbf q''}
}
(C_{\alpha,1_{\nu_{\mathbf q'}}}^{(1)})^*
\langle \Psi_{\alpha,1_{\nu_{\mathbf q'}}}|
\rho_n(\mathbf q)| \Psi_{\beta,1_{\lambda_{\mathbf q''}}}\rangle
C_{\beta,1_{\lambda_{\mathbf q''}}}^{(1)}\\
&=
\sum_{\alpha\beta\nu_{\mathbf q'}
}
(C_{\alpha,1_{\nu_{\mathbf q'}}}^{(1)})^*
(t_{\alpha,n_{\mathbf k}})^*
t_{\beta,n_{\mathbf k}}
C_{\beta,1_{\nu_{\mathbf q'}}}^{(1)}.
\end{align}
Calculation of this quantity can be efficiently performed by
forming the following intermediate
\begin{equation}
D_{n_{\mathbf k},{\nu_{\mathbf q}}}
=
\sum_{\alpha}
C_{\alpha,1_{\nu_{\mathbf q}}}^{(1)}
t_{\alpha, n_{\mathbf k }}.
\end{equation}
With this,
\begin{equation}
\langle \Psi^{(1)}_0
|
\rho_n(\mathbf q)
|\Psi^{(1)}_0\rangle 
=
\sum_{\nu_{\mathbf q}}
(D_{n_{\mathbf k},{\nu_{\mathbf q}}})^*
D_{n_{\mathbf k},{\nu_{\mathbf q}}}.
\end{equation}

\end{document}